\documentclass[conference,a4paper,10pt,twocolumns]{IEEEtran}
\IEEEoverridecommandlockouts

\usepackage{algorithmic,amsmath,amssymb,amsthm,array,bbm,bibentry,cite,color,comment}
\usepackage{enumerate,enumitem,eurosym,float,graphicx,lettrine,mathrsfs,multirow,nomencl}
\usepackage{pict2e,psfrag,ragged2e,relsize,setspace,subfigure,supertabular,tabularx,url,pstricks}
\usepackage[english]{babel}
\usepackage[absolute]{textpos}

\newcommand{\n}{\mathbf{n}}

\newcommand{\p}{\mathbf{p}}

\newcommand{\s}{\mathbf{s}}

\newcommand{\y}{\mathbf{y}}
\newcommand{\z}{\mathbf{z}}

\newcommand{\I}{\mathbf{I}}

\newcommand{\herm}{\mathrm{H}}

\newcommand{\snr}{\mathsf{snr}}
\newcommand{\sinr}{\mathsf{sinr}}
\newcommand{\inr}{\mathsf{inr}}

\DeclareMathOperator*{\ex}{\mathsf{E}}

\newcommand{\bydef}{\stackrel{\cdot}{=}}
\newcommand{\prob}{\mathsf{P}\!}

{		}
{ 		}
{ 		}
{ 		}
{		
	{ 		}
	{		
		{		}
		{		}
		{ 		}
		{ 		}
		{

\newcommand{\DL}{\text{\tiny{DL}}}
\newcommand{\UL}{\text{\tiny{UL}}}
\newcommand{\Tput}{\mathcal{T}_\text{{1MHz}}}

\title{Timely CSI Acquisition Exploiting Full Duplex}
\author{\IEEEauthorblockN{Jes\'{u}s Arnau and Marios Kountouris}
\IEEEauthorblockA{Mathematical and Algorithmic Sciences Lab \\ France Research Center, Huawei Technologies Co., Ltd. \\
						  Email: \{jesus.arnau, marios.kountouris\}@huawei.com}}
\makeindex

\begin{document}
\begin{textblock}{14}(1,.5)
	\begin{center}
		Paper accepted at IEEE WCNC 2017.
	\end{center}
\end{textblock}
\maketitle

\begin{abstract}
In this paper, we propose a method for acquiring accurate and timely channel state information (CSI) by leveraging full-duplex transmission. Specifically, we propose a mobile communication system in which base stations continuously transmit a pilot sequence in the uplink frequency band, while terminals use self-interference cancellation capabilities to obtain CSI at any time. Our proposal outperforms its half-duplex counterpart by at least 50\% in terms of throughput while ensuring the same (or even lower) outage probability. Remarkably, it also outperforms using full duplex for downlink data transmission for low values of downlink bandwidth and received power.
\end{abstract}
%
%\begin{IEEEkeywords}
%_CSIT; CQI aging
%\end{IEEEkeywords}

%=========================================================================
\section{Introduction} \label{sec:intro}
%=========================================================================
In mobile wireless communications, the transmitting end needs to adjust its transmission parameters based on channel information (namely channel state information at the transmitter - CSIT) that may not match the actual conditions at the moment of transmission. For instance, when the user equipment (UE) selects its modulation and coding scheme (MCS), it does so based on the latest feedback from the base station (BS); such feedback often consists of a quantized version of the channel, as estimated from pilot symbols previously transmitted by the UE. Leaving aside the distortion introduced by quantization, or even the noise inherent to any channel estimation process, the delay between estimation and exploitation alone severely impacts the system performance.

To deal with this problem, we take an alternative and novel approach for leveraging full-duplex capabilities at the UE \cite{Choi2010}. Full-duplex radio, that is, transmitting and receiving simultaneously in the same frequency band, is made possible through recent advances in self-interference cancellation (SIC). If self-interference can be suppressed, possibly well below the noise level \cite{Duarte2012,Everett2014}, then simultaneous uplink and downlink transmission could take place, potentially doubling the net throughput. Most of the proposed applications of full duplex have followed this path and have focused on boosting the throughput. 

However, some alternative uses of full duplex have already been proposed in the literature. In \cite{Afifi2013}, it was targeted at enhancing spectrum sensing in a cognitive radio context. In \cite{Han2015}, it was exploited to improve cross-tier intercell interference suppression in an heterogeneous network by allowing pico BSs to simultaneously transmit their desired signal and forward the listened interference. Reference \cite{Du2014} realized the potential of continuous feedback through the full-duplex channel in multiple-input multiple-output (MIMO) communications; the proposal therein substantially reduces the feedback power required to achieve the same multiplexing gains as its half-duplex counterpart. In \cite{Liu2014}, such fast feedback was also exploited to improve AMC in backscatter communications, allowing the transmitter to adapt to a fast-varying channel. More related to our proposal is \cite{Du2015SPAWC}, where full duplex is used to continuously train a BS in open loop and update its precoding matrix.

In this paper, we exploit full duplex in order to provide terminals with timely CSIT. To do so, each BS continuously broadcasts a distinct pilot sequence in the same frequency band used for uplink reception; thanks to their full-duplex capabilities, terminals are able to use the received pilot sequence for estimating the channel at any time. In other words, we exploit full-duplex capabilities to enhance open-loop training. Such training provides timely CSIT that is used to select the most convenient transmission rate.

We explore the potential of our proposed scheme and we compare it with state of the art alternatives. We start by formulating a simplified mathematical model for an uplink where UEs select their transmission rate based on delayed CSIT. We then obtain analytical expressions of throughput and outage probability for different CSIT acquisition schemes. Our numerical results evince that our proposed scheme outperforms half-duplex CSIT acquisition by at least 50\% in terms of throughput while ensuring the same - or even lower - outage probability. Moreover, it can also outperform full-duplex downlink data transmission in cases with low values of downlink bandwidth and received power.

The remainder of the document is structured as follows: Section~\ref{sec:sys_model} describes the system model and the main working assumptions, Section~\ref{sec:performance_analysis} contains the derivations of the main performance metrics, Section~\ref{sec:num_examples} shows numerical examples, and Section~\ref{sec:conclusion} provides some concluding remarks.

%Our solution, sketched in Figure 1, solves the aforementioned problems by having the BS transmit continuously a known sequence, here called beacon, in the same frequency in which it receives data from the UEs; the latter use the beacon to extract CSI at any time they wish, and then autonomously select their MCS according to it. To make this possible, we require both communication ends to be able to work in full-duplex (FD) mode, and in turn to have self-interference mitigation capabilities; in other words, we exploit FD capabilities to enhance open-loop training.
%=========================================================================
\section{System model} \label{sec:sys_model}
%=========================================================================
Consider a mobile cellular communication system where downlink and uplink frequency bands are different, i.e. a frequency division duplex (FDD) system; in fact, our proposal also applies to downlink and uplink decoupling architectures, where UEs can associate to different BSs for each direction \cite{Boccardi2016}. We assume both UE and BS to have a single antenna, even though the proposed idea can be directly extended to MIMO systems.

\begin{figure}%[ht]
	\centering
	\includegraphics[width=.8\columnwidth]{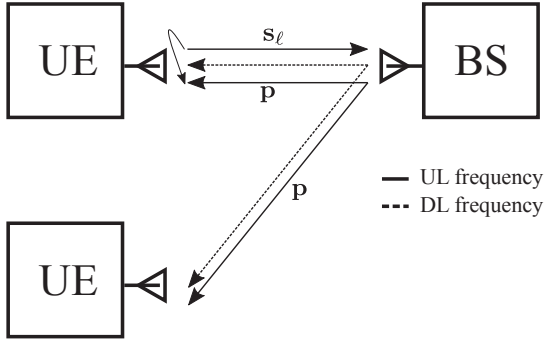}
	\caption{Graphical description of the system model.}
	\label{fig:sys_model}
\end{figure}

We propose that each BS continuously transmits a pilot sequence in the frequency band used for uplink transmissions; see Figure~\ref{fig:sys_model}. UEs with full-duplex radios will be able to estimate their channel based on these pilot sequences, thanks to their SIC capabilities and to the reciprocity of the communication medium. In the following we detail the uplink and downlink signal model, the channel estimation assumed, and the rate selection at the UE.

%We assume that each codeword in the uplink, and each pilot chunk in the downlink, experience a constant channel during their duration, $h_\ell\sim\mathcal{CN}(0,1)$. The autocorrelation between channel samples is given by $\tilde{\rho}\bydef\ex\left[h_\ell h_{\ell+1}^*\right] = J_0(2\pi f_\mathrm{d}T_\mathrm{s})$, where $f_\mathrm{d}$ is the Doppler frequency and $T_\mathrm{s}$ is the time between the samples.

%---------------------------------------
\subsection{Downlink signal model}\label{subsec:dl_smodel}
%---------------------------------------
The $\ell$-th block of signals received by a UE is given by
\begin{equation}\label{eq:signal_model_DL}
\y_{\ell}^\DL = \sum_{j=1}^{N_\mathrm{BS}}\sqrt{\snr_{j}^\DL}\cdot h_{j\ell}\p_j+\sqrt{\inr}\cdot h_0\s_\ell+\n_\ell,
\end{equation}
where $\p_j$ is the pilot sequence transmitted by the $j$-th BS, which consists of $L$ symbols and is received with average signal-to-noise ratio (SNR) $\snr_{j}^\DL$; $\n_\ell$ is the noise vector $\n_\ell\sim\mathcal{CN}(0,\I)$; $h_0$ is the channel remaining after self-interference mitigation, through which the transmitted vector of symbols $\s_\ell$ contaminates the received signal, and $\inr$ is the residual self-interference over noise ratio after cancellation.

We assume $h_0$ to be a Gaussian random variable $h_0\sim\mathcal{CN}(\mu,1)$ \cite{Duarte2012}. However, since the strongest self-interfering paths will be severely attenuated after cancellation (sometimes below the noise level  \cite{Duarte2012,Everett2014}), in the remainder we set $\mu\approx0$. Note that we can emulate different self-interference cancellation capabilities by changing the value of $\inr$.

%---------------------------------------
\subsection{Uplink signal model}\label{subsec:ul_smodel}
%---------------------------------------
When a UE transmits the $\ell$-th block to BS$_j$, the received signal is given by
\begin{equation}
\y_{\ell}^\UL = \sqrt{\snr^\UL}\cdot h_{j\ell} \s_{\ell}+\z_\ell,
\end{equation}
where $\s_\ell$ contains unit-power transmitted symbols, $\y_\ell$ contains the received symbols, and $\z_\ell\sim\mathcal{CN}(0,\I)$. Constant $\snr^\UL$ is the average received SNR, and we will hereafter refer to $\gamma_{\ell}\bydef|h_{\ell}|^2$ as the {\em instantaneous SNR} with a slight abuse of terminology.
As for the channel, we take $h_\ell\sim\mathcal{CN}(0,1)$. The autocorrelation between channel samples is given by $\tilde{\rho}\bydef\ex\left[h_\ell h_{\ell+1}^*\right] = J_0(2\pi f_\mathrm{d}T_\mathrm{s})$, where $f_\mathrm{d}$ is the Doppler frequency and $T_\mathrm{s}$ is the time between the two samples. Note that, differently from UEs, BSs are assumed to have perfect SIC capabilities.

UEs select their transmission rate at the $\ell+1$-th slot (discrete time) based on the latest available channel information, given by
\begin{equation}\label{eq:hat_gamma}
\hat\gamma_\ell = |\hat h_\ell|^2 = |h_\ell+e_\ell|^2.
\end{equation}
We note that, with respect to the true instantaneous SNR $\gamma_{\ell+1}$, the available knowledge $\hat{\gamma_\ell}$ is {\em both delayed and noisy}: delayed because $\gamma_\ell$ and $\gamma_{\ell+1}$ are different (but correlated) as a consequence of the time-varying nature of the channel, and noisy because the underlying channel estimation is affected by an error $e_\ell$. Consistently with conventional data-aided channel estimation, we assume $e_\ell\sim\mathcal{CN}(0,\sigma_\mathrm{e}^2)$; we will elaborate more on this in Section~\ref{subsec:chan_est}. 
%For future reference, we should note that each power sample $|h_\ell|^2$ follows a scaled Chi-squared distribution with two degrees of freedom, $2|h|^2\sim\chi^2(2)$; therefore, the following identities hold: $\ex\left[|h|^2\right] = 1$, $\ex\left[|h|^4\right] = 2.$

In a realistic system, UEs would use their available CSIT to select the most suitable MCS from a finite set; the transmission rate is then determined by the chosen MCS, and an outage occurs if the actual quality of the channel is below the MCS operating threshold.

For tractability, in this work we assume a simplified version of the aforementioned operation. Based on an estimate $\hat\gamma$, the transmitter encodes data at a rate
\begin{equation}\label{eq:rate_def}
R(\hat\gamma) \bydef \log\left(1+\frac{\snr^\UL\cdot\hat\gamma}{\Delta\Gamma}\right),
\end{equation}
where $\Delta$ is a backoff factor emulating the selection of a more protected MCS to guarantee a low packet error rate\footnote{This parameter would in reality be a function of the estimated SNR\cite{Hu2013}, or could be changed over time as the result of an outer loop control \cite{Nakamura2002}, but for our purposes it suffices to consider it fixed.}, and $\Gamma$ represents the SNR gap from Shannon capacity real-world systems exhibit \cite{Mogensen2007} (due to finite constellations, non-ideal channel coding, etc.).

Transmission is successful if the actual $\gamma_{\ell+1}$ is above the one assumed upon encoding. We can write this event as $\mathbbm{1}\!\left[\gamma_{\ell+1}^\mathrm{dB}\geq\hat\gamma_\ell^\mathrm{dB}-\Delta^\mathrm{dB}\right]$ where $\mathbbm{1}[\cdot]$ denotes the indicator function. Then, the outage probability conditioned on an estimation reads as
\begin{equation}
\begin{split}
\epsilon(\hat\gamma) &\bydef 1-\ex\Big[\mathbbm{1}\!\left[\gamma_{\ell+1}^\mathrm{dB}\geq\hat\gamma^\mathrm{dB}-\Delta^\mathrm{dB}\right]\,\Big|\,\hat\gamma_\ell = \hat\gamma\Big] \\
&= \prob\left[\gamma_{\ell+1}\leq\frac{\hat\gamma}{\Delta} \ \middle| \ \hat\gamma_\ell = \hat\gamma\right].
\end{split}
\end{equation}
We further define the effective rate as
\begin{equation}
%\eta(\gamma) = \big(1-\epsilon(\gamma)\big)\cdot\log\left(1+\frac{P_k}{\sigma_\mathrm{z}^2}\frac{\gamma}{\Gamma}\right)
\eta(\hat\gamma) \bydef \big(1-\epsilon(\hat\gamma)\big)R(\hat\gamma)
\end{equation}
and the average effective rate and outage probability, respectively, as
\begin{equation}
\bar\eta \bydef \ex\big[\eta(\hat\gamma)\big], \quad P_\mathrm{out}\bydef \ex\big[\epsilon(\hat\gamma)\big].
\end{equation}

In Section~\ref{sec:performance_analysis}, we obtain analytical expressions for the above metrics, and introduce an additional one to assess the performance of our proposal. Next, we provide some additional explanations on the channel estimation process.

%---------------------------------------
\subsection{Channel estimation}\label{subsec:chan_est}
%---------------------------------------
As explained before, the model adopted in (\ref{eq:hat_gamma}) is consistent with traditional data-aided channel estimation. For the expression of $\sigma_\mathrm{e}^2$ we assume\cite{Ikuno2012}
\begin{equation}
\sigma_\mathrm{e}^2 = \frac{c_\mathrm{e}}{\sinr_\mathrm{csi}}.
\end{equation}
Here, $c_\mathrm{e}$ is a constant that accounts for practical imperfections in the channel estimation process and comprises the training sequence length. As for $\sinr_\mathrm{csi}$, it represents the signal to interference plus noise ratio (SINR) experienced by pilot sequences at the moment of estimation. Let us provide some examples.

%---------------------------------------
\subsubsection{Estimation from uplink signaling}\label{subsec:chan_est_ul}
%---------------------------------------
In the conventional case in FDD, the uplink channel is estimated at the BS from pilot sequences sent by the UEs. Thus, in the absence of interference we simply have $\sinr_\mathrm{csi} =\snr_j^\UL$.

%---------------------------------------
\subsubsection{Full-duplex estimation from downlink pilot sequences}\label{subsec:chan_est_dl}
%---------------------------------------
In our proposal, the channel can be estimated from the pilot signal BSs constantly broadcast thanks to the full-duplex capabilities of the UE. If pilots from different BSs are orthogonal, that is, if $\p_j^\herm\p_l$ is non-zero only when $j=l$, then terminals can distinguish between the channels of different BSs with no interference; for instance, in the case of least-squares channel estimation, we have from (\ref{eq:signal_model_DL}):
%\begin{equation}
%\sqrt{\snr_j^\DL}\widehat{h_{j\ell}} = \frac{1}{L}\p_j^\herm\y_{\ell}^\DL = h_{j\ell}+\frac{\sqrt{\inr}}{L} \p_j^\herm h_0\s_\ell+\frac{1 }{L}\p_j^\herm\n_\ell.
%\end{equation}
\begin{equation}
\p_j^\herm\y_{\ell}^\DL \propto \sqrt{\snr^\DL_j}h_{j\ell}+\sqrt{\inr}\cdot \p_j^\herm h_0\s_\ell+\p_j^\herm\n_\ell,
\end{equation}
and we easily obtain
\begin{equation}\label{eq:sinr_csi_fd}
\sinr_\mathrm{csi} = \frac{\snr_j^\DL}{1+\inr}.
\end{equation}
From (\ref{eq:sinr_csi_fd}) we see that when estimating the channel in full duplex, we incur in a penalty given by the residual self-interference, thus the term $1+\inr$ in the denominator. Our key finding is that, in many cases, the reduction in estimation delay greatly compensates for this noise enhancement. We will demonstrate this in subsequent sections. 

Before doing so, let us recall that $\tilde{\rho}\bydef\ex\left[h_\ell h_{\ell+1}^*\right] = J_0(2\pi f_\mathrm{d}T_\mathrm{s})$, and define the following normalized correlation coefficient:
\begin{equation}\label{eq:norm_rho}
\rho \bydef \frac{\ex\left[\hat h_{\ell}h_{\ell+1}^*\right]}{\sqrt{\ex\left[|\hat h_\ell|^2\right]\ex\left[|h_{\ell+1}|^2\right]}} = \frac{\tilde{\rho}}{\sqrt{1+\sigma_\mathrm{e}^2}}.
\end{equation}
Also recall that, as defined in (\ref{eq:hat_gamma}), $\hat{\gamma}$ is exponentially distributed with mean $\sigma^2 = 1+\sigma_\mathrm{e}^2$.

%=========================================================================
\section{Performance analysis} \label{sec:performance_analysis}
%=========================================================================
\subsection{Outage probability and average effective rate}
We start by obtaining an explicit expression for $\epsilon(\gamma)$, which is given by
\begin{equation}\label{eq:epsilon}
\epsilon(\hat\gamma) = 1-Q_1\left(\sqrt{\frac{2\rho^2}{(1-\rho^2)\sigma^2}\hat\gamma},\sqrt{\frac{2}{(1-\rho^2)\Delta}\hat\gamma}\right),
\end{equation}
where $Q_M(a,b)$ is the Marcum function \cite{Nuttall1975}, $\rho$ is the normalized correlation coefficient (\ref{eq:norm_rho}), and we recall that $\sigma^2 = 1+\sigma_\mathrm{e}^2$; see Appendix~\ref{ap:qfunc} for a detailed explanation. The outage probability $P_\mathrm{out}$ is then given by
\begin{equation}\label{eq:Pout}
\begin{split}
P_\mathrm{out} &= \int_0^\infty\epsilon(x)\frac{1}{\sigma^2}e^{-x/\sigma^2}\,\mathrm{d}x \\
			   &=\frac{1}{2}+\frac{\sigma^2-\Delta}{2\sqrt{(\Delta+\sigma^2)^2-4\tilde\rho^2\Delta}}
\end{split}
%\frac{\sigma^2-\Delta+\sqrt{(\Delta+\sigma^2)^2-4\rho^2\Delta}}{2\sqrt{(\Delta+\sigma^2)^2-4\rho^2\Delta}}
\end{equation}
after plugging (\ref{eq:epsilon}) into the integral and using \cite[Eq. B.48]{Simon2002}.

\begin{figure}%[ht]
	\centering
	\includegraphics{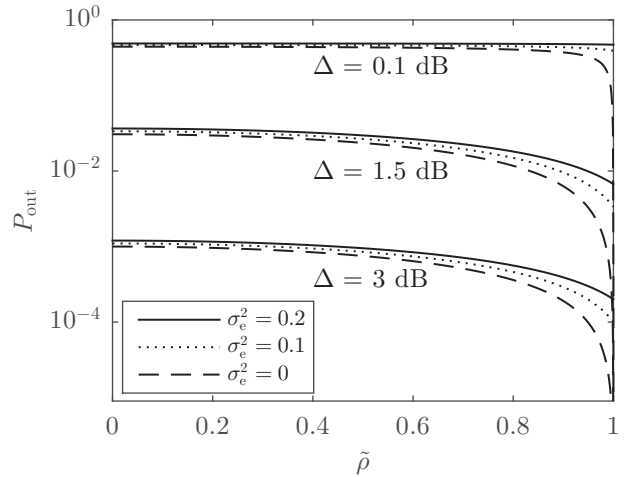}%
	\caption{$P_\mathrm{out}$ as a function of $\tilde\rho$ for different pairs $\{\Delta,\sigma_\mathrm{e}^2\}$.}
	\label{fig:Pout}
\end{figure}
This expression tells us that, in order to make $P_\mathrm{out}$ smaller than 0.5, we need $\Delta>\sigma^2$. Figure~\ref{fig:Pout} numerically evaluates (\ref{eq:Pout}). We can corroborate that increasing $\Delta$ allows us to decrease $P_\mathrm{out}$, but we know from (\ref{eq:rate_def}) that this will decrease the rate; for this reason, increasing $\tilde\rho$ becomes crucial. Our proposed method achieves this by reducing the delay with respect to the last estimated channel value leveraging full duplex.

As for $\bar{\eta}$, we can express it as
\begin{equation}\label{eq:ase}
\begin{split}
\bar\eta &= \int_0^\infty\left(1-\epsilon(x)\right)R(x)\frac{1}{\sigma^2}e^{-x/\sigma^2}\,\mathrm{d}x \\
&=\frac{1}{\sigma^2}\int_0^\infty\!\! Q_1\left(\sqrt{\frac{2\rho^2}{(1-\rho^2)\sigma^2}x},\sqrt{\frac{2}{(1-\rho^2)\Delta}x}\right)e^{-x/\sigma^2} \\
&\times\log\left(1+\frac{\snr\cdot x}{\Delta\Gamma}\right) \,\mathrm{d}x.
\end{split}
\end{equation}

%---------------------------------------
\subsection{Throughput}\label{subsec:thrp}
%---------------------------------------
A fair assessment of our proposal requires taking into account its bandwidth occupancy in the downlink. To do this, let us define the {\em throughput at 1 MHz bandwidth} as
\begin{equation}\label{eq:tput}
\Tput = 1\ \mathrm{MHz}\times\bar\eta\left( \snr_\mathrm{data},\snr_\mathrm{csi},T_\mathrm{csi}\right) \quad \mathrm{Mnats/s}.
\end{equation}
Here, $\snr_\mathrm{data}$ represents the average SNR affecting data decoding (see (\ref{eq:rate_def})), $\snr_\mathrm{csi}$ is the SINR affecting the channel estimation process (see Section~\ref{subsec:chan_est}), and $T_\mathrm{csi}$ is the delay experienced by the CSI, so that $\tilde\rho = J_0(2\pi f_\mathrm{d}T_\mathrm{csi})$.

We must remark that, in defining (\ref{eq:tput}), we are implicitly assuming all power values relative to a bandwidth of 1 MHz. If we scale the bandwidth by a factor $\kappa<1$, we will assume that
\begin{enumerate}
	\item SNR values are divided by $\kappa$ (thus increased). The power budget at the transmitter is the same, but the noise bandwidth is reduced, so that $\snr = \snr_\mathrm{1MHz}/\kappa$.
	\item INR values remain unaffected. If self-interference has a frequency-flat power spectral density around the frequency of operation, then reducing the bandwidth scales both interference and noise power, leaving their quotient unaltered.
\end{enumerate}
In the next section, we numerically evaluate this metric for a number of cases of interest.
%=========================================================================
\section{Numerical Results} \label{sec:num_examples}
%=========================================================================
In this section we evaluate (\ref{eq:Pout}) in order to plot $P_\mathrm{out}$, and (\ref{eq:ase}) through numerical integration to plot $\Tput$. We compare our proposal to three benchmark curves described below.
%---------------------------------------
\subsection{Description of the curves}\label{subsec:curves}
%---------------------------------------

%\begin{itemize}
	% PeRFECT CSI
	\paragraph  {Perfect CSI (PCSI)} As a half-duplex upper bound we consider the case of having perfect CSIT. In our model, this translates into unlimited acquisition SNR and zero delay:
	\begin{equation}
\Tput^\mathrm{PCSI} = 1\ \mathrm{MHz}\times\bar\eta\left( \snr^\UL,\infty,0\right).
	\end{equation}
	The outage probability is by definition zero.

	% Probing.
	\paragraph {Probing (PROBE)} As a half-duplex baseline we consider the case of the UE probing the channel before transmission; there is no self-interference penalty when estimating the channel, but the delay $T_\mathrm{pr}$ equals at least the round-trip delay, i.e.
	\begin{equation}
	\Tput^\mathrm{PROBE} = 1\ \mathrm{MHz}\times\bar\eta\left( \snr^\UL,\snr^\UL,T_\mathrm{pr}\right).
	\end{equation}
	
	% FuLL DUPLEX CSI
	\paragraph {Full-duplex CSI acquisition (FDCSI)} Our solution, as previously described: channel estimation is affected by residual self-interference, but delay is minimal $T_\mathrm{min}<<T_\mathrm{pr}$:
	\begin{equation}
	\Tput^\mathrm{FDCSI} = 1\ \mathrm{MHz}\times\bar\eta\left( \snr^\UL,\frac{\snr^\DL}{\kappa(\inr+1)},T_\mathrm{min}\right).
	\end{equation}

	% FuLL DUPLEX DATA
	\paragraph {Full duplex for data transmission (FDDATA)} For comparison purposes, we evaluate the throughput of an alternative full-duplex solution that uses self-interference cancellation capabilities to receive data. In this case we add up the throughput of both uplink (first term) and downlink (second term), assuming they both use probing to obtain their CSI:
	\begin{equation}
	\begin{split}
	\Tput^\mathrm{FDDATA} &= 1\ \mathrm{MHz}\times\bar\eta\left( \snr^\UL,\snr^\UL,T_\mathrm{pr}\right) \\
						 &+ \kappa\ \mathrm{MHz}\times\bar\eta\left(\frac{\snr^\DL}{\kappa(\inr+1)},\frac{\snr^\DL}{\kappa(\inr+1)},T_\mathrm{pr}\right).
	\end{split}
	\end{equation}
	Note that we neglect interference from other BSs in the downlink.
%\end{itemize}
%=========================================================================
\subsection{Examples} \label{subsec:examples}
%=========================================================================

In all the examples that follow we have set $f=2$~GHz, $\Gamma = 1$~dB, $\mathsf{snr}^\UL = 5$~dB, $T_\mathrm{min} = 2$~ms, $T_\mathrm{pr} = 4$~ms, $\Delta_\mathrm{PROBE}=\Delta_\mathrm{FDDATA} = 5.6$~dB. All power values are relative to 1~MHz bandwidth as explained in Section~\ref{subsec:thrp}, and all the FDCSI curves have been obtained with $\kappa=1/10$; the latter means we only use 100 kHz to send the downlink pilots\footnote{Some simple computations with the autocorrelation function of the channel show that, with a bandwidth of 100 kHz, more than 150 pilot symbols can be transmitted within the 80\% coherence of the channel at a speed of 50 km/h, $f_\mathrm{c} = 2$~GHz.}. Constant $c_\mathrm{e}$ is set to 0.0544 \cite{Ikuno2012}.

\begin{figure*}%[ht]
\centering
\includegraphics{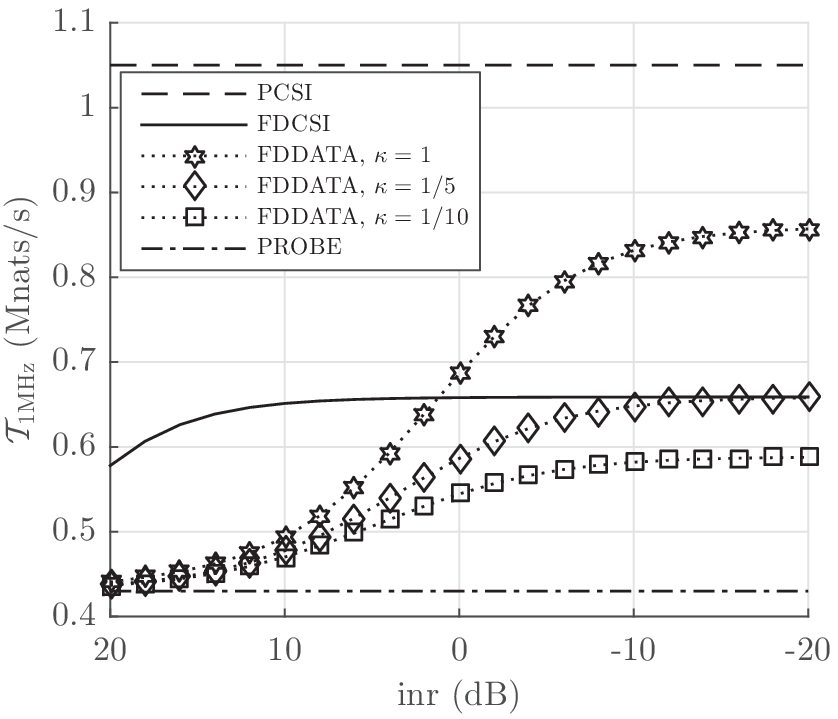}%
\includegraphics{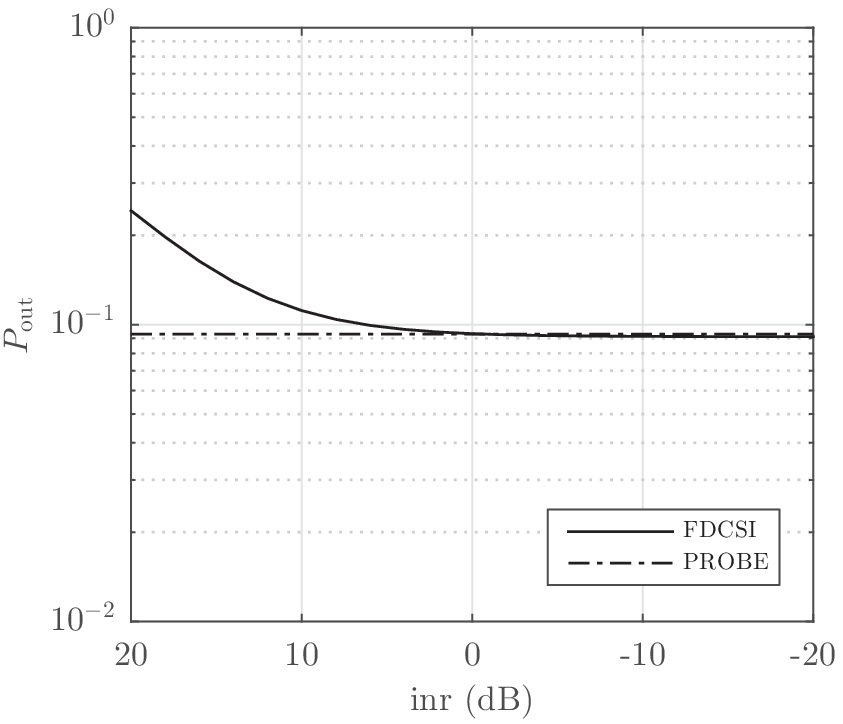}
\caption{$\Tput$ (left) and $P_\mathrm{out}$ (right) as a function of $\mathsf{inr}$, $\snr^\DL = 5$ dB, UE speed 15 km/h.}
\label{fig:fig_vs_inr_5dB}
\end{figure*}

Figure~\ref{fig:fig_vs_inr_5dB} plots the evolution of the desired performance metrics with respect to the $\inr$. In this example we have set $\snr^\DL = 5$~dB, the UE speed to 15~km/h, and $\Delta_\mathrm{FDCSI} = 3$~dB; the latter has been chosen based on (\ref{eq:Pout}) to ensure the same asymptotic $P_\mathrm{out}$ as with probing. Focusing on FDCSI, we see that its throughput quickly becomes constant with respect to the residual $\inr$, roughly after $\inr=0$~dB; moreover, it outperforms PROBE by 53\% in terms of throughput, and achieves the same performance as FDDATA when the latter occupies twice as much bandwidth in the downlink ($\kappa = 1/5$).

\begin{figure*}%[ht]
	\centering
	\includegraphics{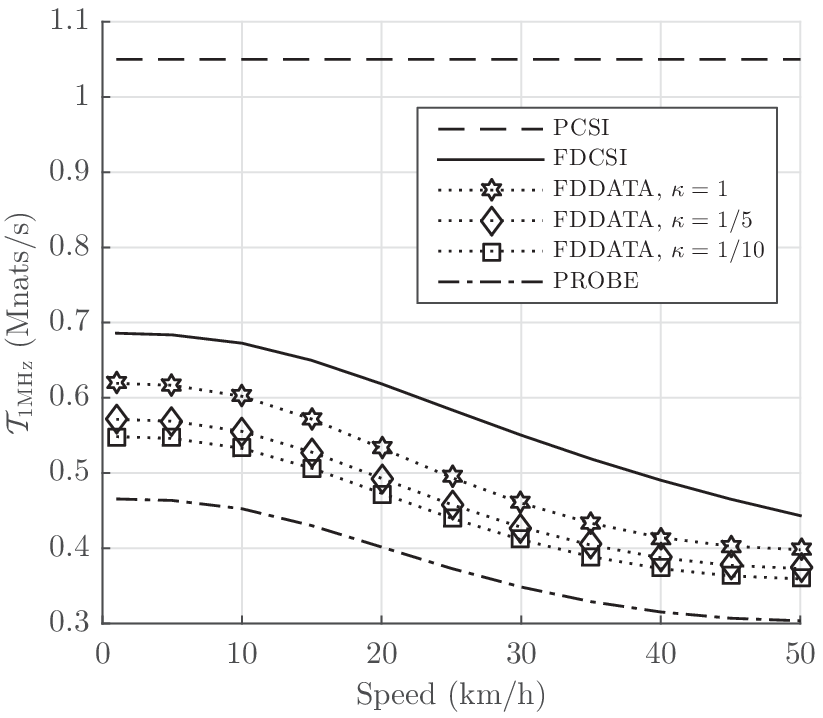}%
	\includegraphics{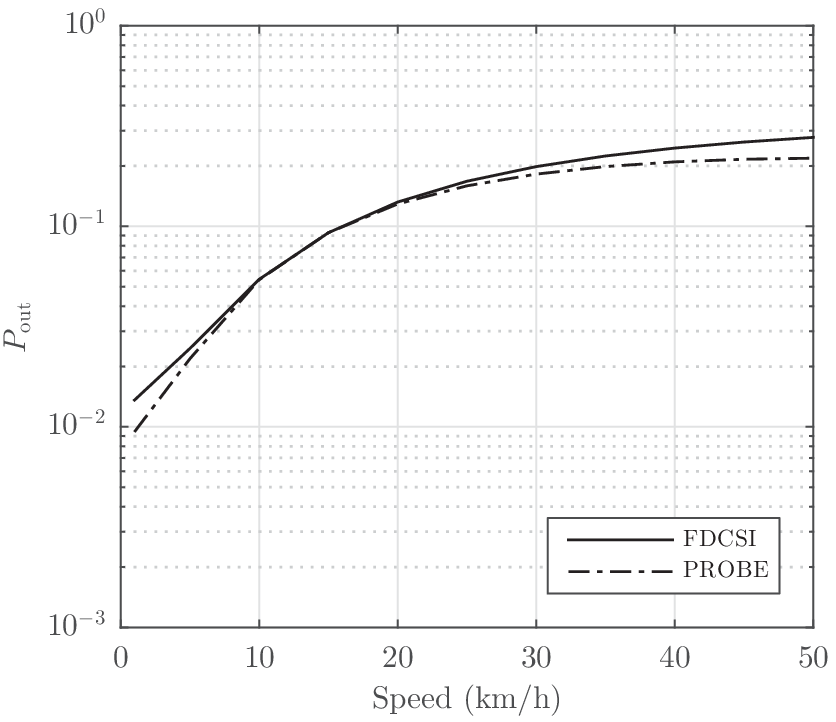}
	\caption{$\Tput$ (left) and $P_\mathrm{out}$ (right) as a function of speed. $\snr^\DL = 0$ dB, $\mathsf{inr} = -5$ dB.}
	\label{fig:T_vs_speed_0dB}
\end{figure*}

Figure~\ref{fig:T_vs_speed_0dB} illustrates the case of changing UE speed with $\mathsf{inr} = -5$~dB, $\Delta_\mathrm{FDCSI}=3.1$~dB and lower downlink SNR, $\snr^\DL = 0$~dB. We can see that the throughput gap with respect to PROBE is roughly the same as in Figure~\ref{fig:fig_vs_inr_5dB}. But, remarkably, FDCSI outperforms now FDDATA for all values of $\kappa$.  
%=========================================================================
\subsection{Discussion} \label{sec:discussion}
%=========================================================================
Through the following items we highlight relevant details in the performance results and point out to some caveats.
\subsubsection{Downlink power and bandwidth}
Figure~\ref{fig:fig_vs_inr_5dB} and Figure~\ref{fig:T_vs_speed_0dB} suggest that our proposal is more bandwidth efficient than using full duplex for data transmission when the received downlink power is low. Conversely, they show that, when larger bandwidth and high powers are available in the downlink, full duplex for data makes a much more efficient use of them.

The observation above has been made taking into account the throughput of both links, but this might not be the best metric in certain relevant scenarios. If we exclusively focus on improving the uplink, FDCSI seems to be always a better choice.

\subsubsection{Unknown interference}
In this paper we have focused on tracking the evolution of the channel over time and have disregarded the effect of time-varying interference. The underlying assumption is that interference events come from within the same cell, and that, when they happen, they can be regarded as collisions that will be dealt with by the MAC layer.

If there is out-of-cell time-varying interference, our solution FDCSI cannot track it directly, neither can the baseline PROBE unless interference changes very slowly over time. Both alternatives would probably need to resort to larger time-varying backoff values.%, which in principle would have to be time-varying.

%\subsubsection{Comparison with TDD-based solutions}

%=========================================================================
\section{Conclusions} \label{sec:conclusion}
%=========================================================================

We have proposed a mobile communication system where a BS continuously transmits a pilot sequence in the uplink frequency band, so that mobile terminals can acquire timely CSIT by leveraging full duplex and self-interference cancellation capabilities. In all relevant evaluation scenarios considered here, our proposal outperforms its half-duplex counterpart by at least 50\% in terms of throughput while ensuring the same (or even lower) outage probability. Interestingly, in cases with low values of downlink bandwidth and received power, it can also outperform full-duplex radios used for downlink data transmission.

\appendices
%=========================================================================
\section{Derivation of $\epsilon(\gamma)$} \label{ap:qfunc}
%=========================================================================

Both $\gamma_{\ell+1}$ and $\hat\gamma_\ell$ are jointly exponentially distributed; equivalently, each of their doubles is chi-squared distributed with two degrees of freedom. In consequence, we have that \cite[Eq. 3.17]{Simon2002}
\begin{equation}
\begin{split}
f_{\gamma_{\ell+1},\hat\gamma_\ell}(u,v) &= \frac{1}{\sigma^2(1-\rho^2)}\exp\left(-\frac{1}{1-\rho^2}\left(u+\frac{v}{\sigma^2}\right)\right) \\
	&\times I_0\left(2\frac{|\rho|\sqrt{uv}}{(1-\rho^2)\sigma}\right)
\end{split}
\end{equation}
after applying a simple change of variables. Note that $\rho$ is the correlation coefficient between the underlying Gaussian random variables, $h_{\ell+1}$ and $h_\ell+e_\ell$.

The probability distribution of $\gamma_{\ell+1}$ conditioned on the last estimation $\hat\gamma_\ell$ is now given by
\begin{equation}\label{eq:cond_dist}
\begin{split}
f_{\gamma_{\ell+1} | \hat\gamma_\ell}(u,v) &= \frac{f_{\gamma_{\ell+1},\hat\gamma_\ell}(u,v)}{f_{\hat{\gamma_\ell}}(v)} \\
					   &= \frac{1}{1-\rho^2}\exp\left(-\frac{1}{1-\rho^2}\left(u+\frac{v}{\sigma^2}\right)+\frac{v}{\sigma^2}\right) \\
&\times I_0\left(2\frac{|\rho|\sqrt{uv}}{(1-\rho^2)\sigma}\right),
\end{split}
\end{equation}
which is the pdf of a scaled non-central chi-squared distribution with two degrees of freedom. %the expression of $\epsilon(\gamma)$ thus follows after applying \cite[Eq. 2.45]{Simon2002} and an appropriate change of variables.
The derivation continues as follows:
\begin{equation}\label{eq:deriv_epsilon}
\begin{split}
\epsilon(x) &= \prob\left[\gamma_{\ell+1}\leq\frac{x}{\Delta} \ \middle| \ \hat\gamma_\ell = x\right] \\
&= 1-\int_{x/\Delta}^\infty f_{\gamma_{\ell+1} | \hat\gamma_\ell}(t,x)\,\mathrm{d}t \\
	   &= 1-\int_{x/\Delta}^\infty \frac{f_{\gamma_{\ell+1},\hat\gamma_\ell}(t,x)}{f_{\hat\gamma_\ell}(x)}\,\mathrm{d}t.
\end{split}
\end{equation}
Before plugging (\ref{eq:cond_dist}) into (\ref{eq:deriv_epsilon}) we apply the following change of variables, $z = \sqrt{2t/(1-\rho^2)}$, $\mathrm{d}t = (1-\rho^2)z\,\mathrm{d}z$, so that
\begin{equation}\label{eq:dfd}
\begin{split}
\epsilon(x) &=1- \\ 
				&\int_c^\infty \!\! z\exp\left(-\frac{z^2+\frac{2\rho^2x}{(1-\rho^2)\sigma^2}}{2}\right)I_0\left(\sqrt{\frac{2\rho^2x}{(1-\rho^2)\sigma^2}}z\right)\,\mathrm{d}z \\
                        &=1-Q_1\left(\sqrt{\frac{2\rho^2}{(1-\rho^2)\sigma^2}x},\sqrt{\frac{2}{(1-\rho^2)\Delta}x}\right),
\end{split}
\end{equation}
where $c = {\sqrt{\frac{2x}{(1-\rho^2)\Delta}}}$, and the last equality follows from the definition of the $Q_M$ function \cite{Nuttall1975}.
%\addcontentsline{toc}{chapter}{References}
\bibliographystyle{IEEEtran}
\bibliography{IEEEabrv,journals,books_and_others,confs}
%=============================================
\end{document}